\documentclass[twocolumn,preprintnumbers,amsmath,amssymb,pra]{revtex4}
\usepackage{amsmath,amssymb,graphicx,epstopdf,bm,soul,upgreek, hyperref,natbib}
\usepackage{graphicx}
\usepackage{color}
\definecolor{OliveGreen}{rgb}{0,0.6,0}
\usepackage{dcolumn}
\usepackage{bm}
\usepackage{amssymb,float}
\usepackage{cancel}
\usepackage{ulem}

\begin{document}

\title{Coupling between exciton-polariton corner modes mediated through edge states}%

\author{R. Banerjee}\email[Corresponding author:~]{rimi001@e.ntu.edu.sg}
\author{S. Mandal}
\author{T.C.H. Liew}\email[Corresponding author:~]{tchliew@gmail.com}

\affiliation{Division of Physics and Applied Physics, School of Physical and Mathematical Sciences, Nanyang Technological University, Singapore 637371, Singapore}


\begin{abstract}

Recently realized higher order topological insulators have taken a surge of interest among the theoretical and experimental condensed matter community. The two dimensional second order topological insulators give rise to zero dimensional localized corner modes that reside within the band gap of the system along with edge modes that inhabit a band edge next to bulk modes. Thanks to the topological nature, information can be trapped at the corners of these systems which will be unhampered even in the presence of disorder. Being localized at the corners, the exchange of information among the corner states is an issue. Here we show that the nonlinearity in an exciton polariton system can allow the coupling between the different corners through the edge states based on optical parametric scattering, realizing a system of multiple connectible topological modes.
\end{abstract}

\maketitle
{\textit{Introduction}---}  Topological insulators (TIs) have attracted attention in the past decade due to their unique exotic property, namely the appearence of backscattering-immune edge states, which can propagate against perturbation without being backscattered. TIs have been explored in various systems including electronics~\cite{HasanKane2010,QiZhang2011},  photonics~\cite{Haldane2008,WangChong2008,HafeziDemler2011,WangChong2009,HafeziMittal2013,PooWu2011,
RechtsmanZeuner2013,ChenJiangJiang}, cold atoms~\cite{Jotzu2014,Aidelsburger2015}, exciton-polaritons~\cite{BardynKarzig2015,NalitovSolnyshkov2015,KarzigBardyn2015, KlembtHarder2018, YaroslavKartashov2017, BanerjeeLiew2018}, acoustics~\cite{Yang2015,FleuryKhanikaev2016}, etc. Recently the concept of topological phases was extended to higher order topological phases \cite{BenalcazarBernevig2017,BenalcazarBernevig2017_2,SongFang2017,LangbehnPeng2017,Schindler2018,
GarciaPeri2018,ImhofBerger2018,PetersonBenalcazar2018,Ezawa2018,XieWang2018,NohBenalcazar2018,
XueYang2019,Geier2018} that go beyond the conventional bulk-boundary correspondence~\cite{HasanKane2010}. A two-dimensional  second order topological insulator can host topologically protected zero dimensional  gapless corner states along with one dimensional gapped edge states. The zero dimensional corner states have been realized using quantization of quadrupole moments in square lattices~\cite{BenalcazarBernevig2017,BenalcazarBernevig2017_2}, classical mechanical systems~\cite{GarciaPeri2018}, electromagnetic metamaterials~\cite{ImhofBerger2018,PetersonBenalcazar2018}, breathing kagome lattices~\cite{Exawa2018,KunstMiert2018,ArakiMizoguchi2018}, and acoustic metamaterials~\cite{XueYang2019,Xiangweiner2019}. Due to the topological properties of these corner states, information can be trapped at the corners of the system, which will be unhampered even in the presence of disorder, making it a potential candidate for information processing~\cite{XieWang2018,LiWang2018,RedondoBell2018,Tambasco2018,WangPang2018,HassanKunst2019,
ChenDeng2018,SusstrunkHuber2015,NashKleckner2015,YangJia2019}. But corner states, as with other topological modes, are well isolated from each other even in the presence of disorder making it difficult for them to overlap (they are orthogonal eigenstates). Consequently, it is far from obvious whether there are ways in which different corner states can interact. Although information processing necessitates operating with a coupling of multiple modes, the coupling of multiple topological modes is unexplored in the literature (especially for corner states). \par
In this letter we consider theoretically an array of coupled exciton-polariton micropillars arranged in a square lattice. Exciton-polaritons are  hybrid light matter quasiparticles that arise from the strong coupling of quantum well excitons and microcavity photons. They are well-known for a variety of nonlinear effects, typically studied in planar microcavities \cite{DengHaug2010,CarusottoCiuti2013}. Several experiments
\begin{figure}[t]
\centering
\includegraphics[width=\linewidth]{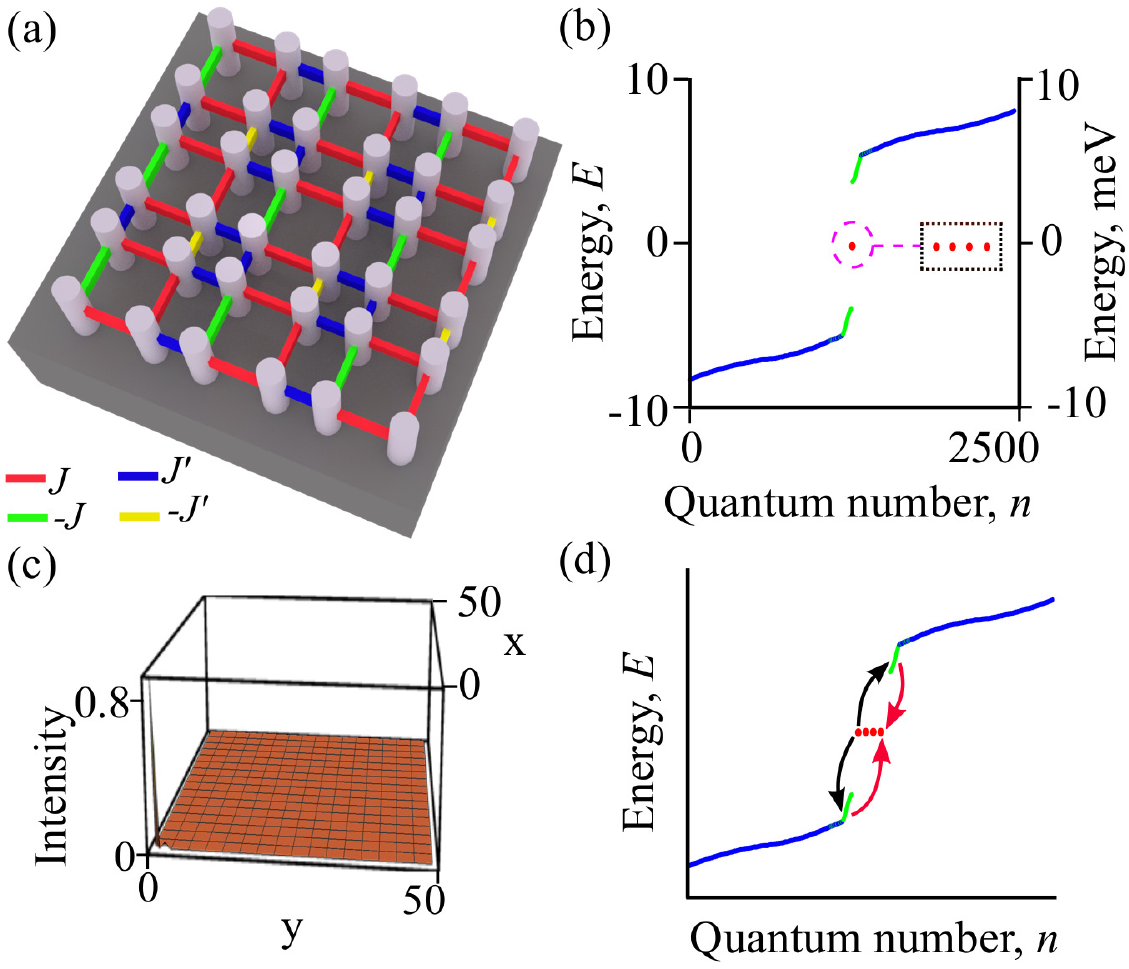}
\caption{(a) Schematic diagram of a square lattice formed by coupled exciton-polariton micropillars with four different hoppings, $J$, $-J$, $J'$, and $-J'$, indicated by different colours. (b) Energy eigen-values of the system consisting of $50 \times 50$ micropillars, as a function of the quantum number $n$. The modes corresponding to $n=1249-1252$ are the corner states appearing at $E=0$, denoted by red. The bulk and edge states are shown in blue and green respectively. (c) Spatial profile of the pumped corner state corresponding to $n=1251$. (d) Schematic diagram of the optical parametric scattering processes in the system. First polaritons from the pumped corner state scatter to edge states  (black arrows) and then the polaritons from the edge states scatter back to the adjacent corner state (red arrows). Parameters:  $J^{\prime}=5$, $\Delta=0$.}
\label{fig:sch_model}
\end{figure}
were accomplished with arrays of coupled micropillars, where the hopping between two micropillars is realized by having overlap between them \cite{Rodriguez2016,Milicevic2015,Sala2015,Winkler2016,Klembt2017, ZhangBrod2015}. These systems have allowed the implementation \cite{KlembtHarder2018} of schemes \cite{BardynKarzig2015,NalitovSolnyshkov2015,KarzigBardyn2015} for first order topological bandstructures. Theoretical works have studied nonlinear effects in such systems, including: inversion of topology \cite{BleuSolnyshkov2016}, formation of solitons \cite{KartashovSkryabin2016,GulevichShelykh2017,LiYe2018}, antichiral behavior\cite{Mandal2019}, and bistability \cite{YaroslavKartashov2017}.

Here we consider a second order topological polariton bandstructure, which is based on achieving hopping between sites with opposite sign. This can not be achieved just by varying the overlap between neighbouring micropillars, but can be implemented by placing auxiliary micropillars between a “main” lattice of micropillars. This follows a generic scheme introduced in Ref. \cite{Keil_2016} for tight-binding lattices, which we verified starting from a particular polariton potential profile. Having established polariton corner states, we study the influence of polariton-polariton scattering, which allows them to couple to edge states. It is via edge states that polariton corner states can interact, where the excitation of one corner state causes excitation of its neighbor, which would not be possible in the linear regime. We have demonstrated transfer of information encoded in a binary state from one corner to the next, which can occur even in the presence of a realistic level of disorder. We find that the mechanism of information transfer proceeds both faster and with lower required power than the same mechanism considered in a regular non-topological square lattice.\par


{\textit{Scheme---}} We consider a square lattice of coupled exciton-polariton micropillars as shown in Fig.~\ref{fig:sch_model}(a). For simplicity, we neglect the spin degree of freedom in the system and consider a single mode of each micropillar, which evolves according to the driven-dissipative nonlinear Schr\"odinger equation
\begin{align}
i\hbar\frac{\partial\psi_{i}}{\partial t}&=\left(\Delta-\frac{i\Gamma}{2}+iP\right)\psi_{i}+\sum_{\langle j\rangle} J_{ij} \psi_{j}\notag\\
&+\alpha|\psi_{i}|^2\psi_{i}-i\alpha_{NL}|\psi_{i}|^2\psi_{i}+F_i
\label{eq:GPEq1st}
\end{align}
where $\Delta$ is the energy detuning between the polariton mode energy (onsite energy) and the laser energy, and  $\Gamma$ is the polariton dissipation. $P$ is a nonresonant pump applied uniformly to all micropillars in the system and as a result the nonlinear loss term $\alpha_{NL}$ is inevitable. $\alpha$ is the strength of nonlinear interaction and  $F$ is a coherent driving field (i.e., laser). Next we move to the dimensionless units by making the following transformations: $t \rightarrow {t \hbar}/{J} $ and $\psi_i \rightarrow \psi_i \sqrt{(J/\alpha)} $, where $J$  is the weakest hopping amplitude. With these  choices Eq.~(\ref{eq:GPEq1st}) becomes
\begin{align}
i\frac{\partial\psi_{i}}{\partial t}&=\left(\Delta-\frac{i\Gamma}{2}+iP\right)\psi_{i}+\sum_{\langle j\rangle}J_{ij}\psi_{j}\notag\\
&+\left(1-i\alpha_{NL}\right)|\psi_{i}|^2\psi_{i}+F_i
\label{eq:GP1}
\end{align}
where the polariton-polariton nonlinear interaction is scaled to unity. All the energy terms in the equation are normalized by a factor $J$, $\alpha_{NL}$ is normalized by  $\alpha$, and $F_i \rightarrow  F_i\sqrt{(\alpha/J^{3})}$. We consider four different hopping terms ($J, -J,J^{\prime}$, and -$J^{\prime}$)  realizing the potential described in Fig.~\ref{fig:sch_model}(a). In writing Eq.~(\ref{eq:GP1}), we have assumed a tight-binding approximation. Exciton-polaritons can also be modelled directly from a continuous model, discussed in the supplementary material (SM). There, we also explain how hopping terms of opposite sign can be achieved, by making use of auxilliary micropillars in the lattice. Although we operate with dimensionless units, we also provide typical real units, corresponding to setting the coupling $J=1$ meV (corresponding to Ref. [\!\citenum{Vasconcellos_2011}]). \par
Neglecting at first the excitation, decay, and nonlinear terms, the energy spectrum ($E$) as a function of eigenstate quantum number ($n$)  is given in  Fig.~\ref{fig:sch_model}(b). The zero energy modes correspond to the corner states, which are well separated in energy from other modes of the system. This should be expected as, in the absence of nonlinear terms, the model is essentially the same as that applicable to coupled microwave resonators previously shown to exhibit the same topological corner states \cite{PetersonBenalcazar2018}. Our main aim here 
is to excite one of the corner modes coherently and couple to other corner modes without affecting the topological property of the system. In this way topological corner modes may be used to store information as well as support exchange of information among themselves. To do this we consider all terms as described in  Eq.~(\ref{eq:GP1}) and take $F_{i}$
as proportional to the amplitudes in each micropillar corresponding to one of the corner states, i.e., $F_{i}=f_{s} \psi _{i}^{(c)}$, where $\psi^{(c)}$ is the corner state eigenfunction. We choose the pump profile with the same spatial profile as the eigen state corresponding to $n=1251$, shown in Fig. \ref{fig:sch_model}(c), and also for simplicity we fix ${\Gamma}/{2}=P$. With proper choice of parameters, we consider parametric instability in the system where pairs of polaritons from this corner state can scatter to the edge modes while conserving energy. This regime of optical parametric oscillation (OPO) was previously deeply studied in planar microcavities~\cite{CiutiSchwendimann2003,CiutiSchwendimann2001,Whittaker2001,Stevenson2000,CiutiCarusotto2005}. Due to secondary parametric scattering processes ~\cite{SavvidisCiuti2001,TartakovskiiKrizhanovskii2002}, we expect that the edge states will couple to another corner. In this way with the help of the edge states we can nonlinearly couple the two corners, as described by the schematic figure \ref{fig:sch_model}(d). \par   

\begin{figure}[htbp]
\centering
\includegraphics[width=\linewidth]{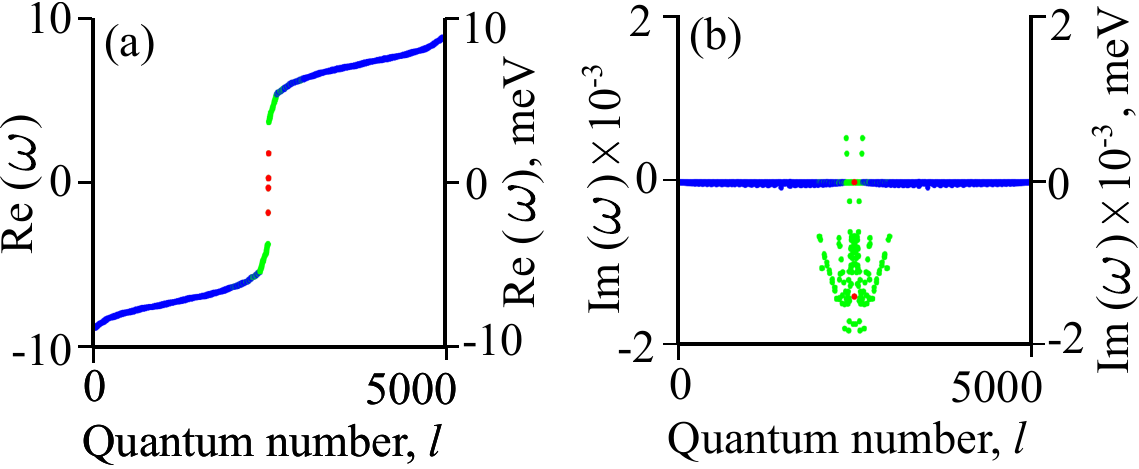}
\caption{Real  and imaginary parts of the eigenvalues of the fluctuations as a function of quantum number $l$ in (a) and (b) respectively. The states located at the corners are shown in red. The bulk and the edge states are represented by blue and green dots respectively. Note that the total eight states corresponding to the four corners are in the band gap in (a). The positive imaginary part implies instability in the system and in the real space those four states having Im$(\omega)>0$ correspond to different edge states. This also indicates that the pumped corner state couples to more than one edge state. Parameters: $J^{\prime}=5$, $\Delta=-0.3$, $\alpha_{NL}=0.2$, $f_{s}=\sqrt{1.4}$.}
\label{fig:bogo_fluc}
\end{figure}

{\textit{Parametric Instability---}}
To investigate parametric instability we first drive the system to a steady state, which is obtained by solving Eq.~(\ref{eq:GP1}), and then study the behaviour of linear (Bogoliubov) fluctuations 
\begin{align}
{\psi_{i}}=\psi_{(0,i)}+u_{i}e^{-i\omega t}+v_{i}^{*}e^{i\omega^{*} t}
\label{eq:2}
\end{align} 
 $\psi_0$ is the stationary solution of Eq.~(\ref{eq:GP1}), which essentially takes the form of the driven corner state; $u$ and $v$ are spatial functions of the fluctuations. $\psi_{(0,i)}$ is value of the $\psi_0$ at the lattice site $i$. Similarly $u_{i}$ and $v_{i}$ are the amplitudes of fluctuations at lattice site $i$. Here  $\omega$ is the frequency of the fluctuations, which is in general complex to encapsulate the instabilities of the system. Substituting Eq.~(\ref{eq:2}) into Eq.~(\ref{eq:GP1}), we obtain the following eigenvalue equations 
\begin{align}
\omega u_i&=\left(iP'+\Delta'\right) u_{i}+\sum_{\langle j\rangle}J_{ij} u_{j}+\left(1-i\alpha_{NL}\right)\psi_{(0,i)}^{2} v_{i}\notag\\
\omega v_i&=\left(iP'-\Delta'\right) v_{i}-\sum_{\langle j\rangle}J_{ij} v_{j}-\left(1+i\alpha_{NL}\right)({\psi^{*}_{(0,i)}})^{2} u_{i}
\label{eq:3}
\end{align} 
where $\Delta'=\Delta+2|\psi_{(0,i)}|^2$ and $P'=P-\frac{\Gamma}{2}-2\alpha_{NL}|\psi_{(0,i)}|^2$.
The eigenvalues of Eq.~(\ref{eq:3}) are plotted  as a function of the quantum number $l$ of the fluctuation in Fig.~\ref{fig:bogo_fluc}. The modes corresponding to Im$(\omega)>0$ indicate instability in the system and these modes correspond to edge states when plotted in real space (see the SM for their spatial profiles). Thus, polariton-polariton scattering induces coupling between the corner mode and edge modes. \par

\begin{figure}[t]
\centering
\includegraphics[width=\linewidth]{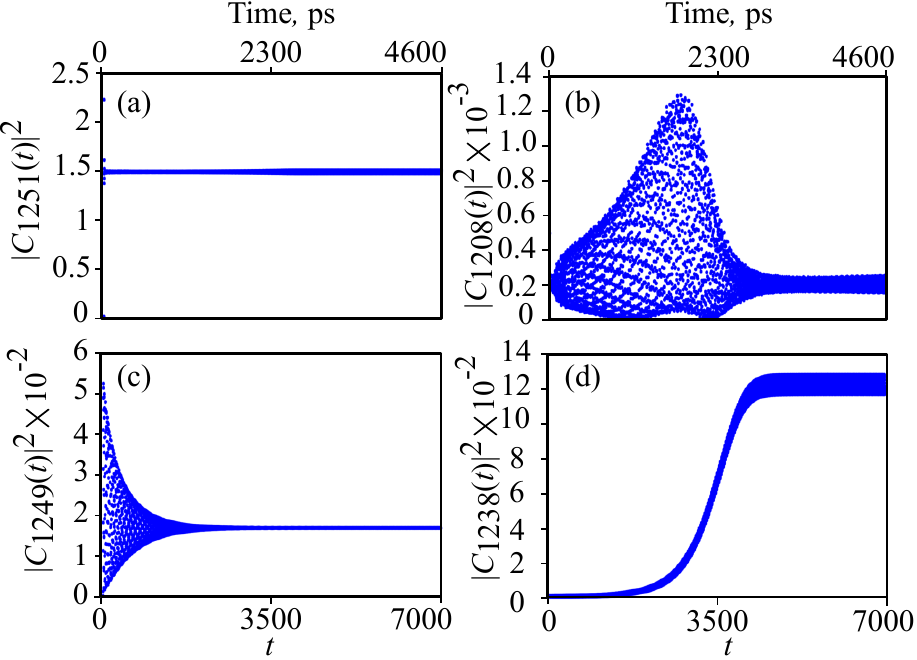}
\caption{The overlap $|C_{n}(t)|{^2}$ as a function of time for (a) the excited corner, $n=1251$, (b) the edge state corresponding to $n=1208$, and (c) the adjacent corner, $n=1249$. (d) In later time, due to higher order terms, another edge state corresponding to $n=1238$ appears in the system. The nonzero value of intensity of the adjacent corner indicates that both the corners are coupled.}
\label{fig:cn_sqr}
\end{figure}
{\textit{Corner-corner coupling mediated via parametric interaction---}}
Now instead of depending on the linear Bogoliubov theory we directly simulate the dynamics of the system described by Eq.~(\ref{eq:GP1}) starting from an initial vacuum state corresponding to zero mean-field. At each time step, the solution can be expanded as a linear superposition of the eigenstates of the linear system as
\begin{align}
\psi_{i}(t)=\sum_{n}C_{n}(t)\psi_{(n,i)}
\label{eq:4}
\end{align}
where $C_{n}(t)=\sum_{i}\psi_{i}(t) \psi_{(n,i)}^{*}$. Physically $|C_{n}(t)|{^2}$ represents the intensity (or overlap) of each eigenstate. A plot of $|C_{n}(t)|{^2}$ as a function of time is shown in  Fig.~\ref{fig:cn_sqr} for $n$ corresponding to the excited corner,  edge states and another adjacent corner state.  Fig.~\ref{fig:cn_sqr}(a) shows that in a very short time (much faster than the chosen range in the time plotted), the excited corner reaches its steady state and  starts to couple to the edge states as shown in Fig.~\ref{fig:cn_sqr}(b). Note that the excited corner state couples with more than one edge state and they have different intensity profiles with time (here as an example we have plotted only one). In Fig.~\ref{fig:cn_sqr}(c) the intensity profile of the adjacent corner is plotted with time and the nonzero value indicates that there is  coupling between the two corners. Without the nonlinear terms, the coupling of the excited corner state to the edge states or its adjacent corner state vanishes (see the SM).  In a later time we observe significant intensity corresponding to another edge state as shown in Fig~\ref{fig:cn_sqr}(d), which eventually  has no effect on the steady adjacent corner state. \par
The possibility that the nonlinear terms directly couple the corner states can not be discarded from the obtained results so far. However the analysis of the linear fluctuations can be repeated in the eigenbasis. Doing this (in the SM) we find that there is no parametric instability of a corner state into another corner state but only into the edge states. Thus, we ascertain that polaritons from the pumped corner mode couple first to an adjacent edge mode and it is via this edge mode that coupling to the adjacent corner mode is achieved. 


The most important parameters in our scheme to realise the coupling are the nonlinear interaction and loss terms. Here the nonlinear self-energy in the system becomes about $1.36$ meV, which is within experimental limits \cite{Sun_2017}. We have taken the nonlinear loss coefficient as $\alpha_{NL}=0.2$, where a similar value was used in ~Ref. \cite{KeelingBerloff2008}. \par

 {\textit{Demonstration of transfer of binary information.}---} Here we demonstrate that the coupling between corners mediated by parametric instability is sufficient to transfer binary information. Such demonstration is based on using near-resonant coherent laser fields at each corner to place them in a bistable regime, which forces each corner state to either be in a low or high intensity state. Switching the state of one corner results in a later switch of the adjacent corner state, corresponding to a transfer of information. Remarkably, this can occur even in the presence of a realistic level of disorder.\par

\begin{figure}[htbp]
\centering
\includegraphics[width=\linewidth]{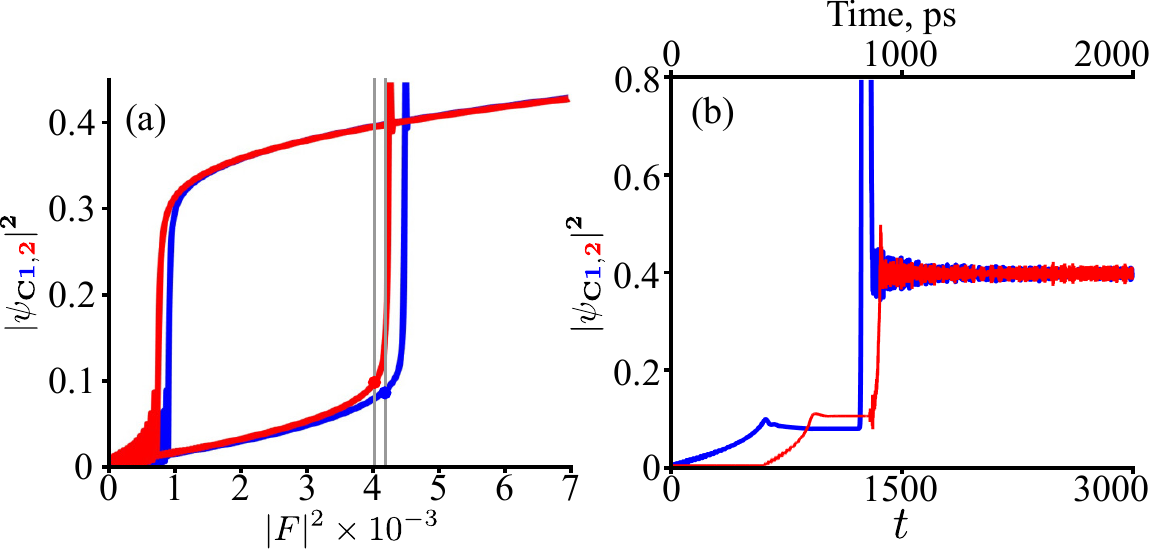}
\caption{(a) Hysteresis curve of polariton density vs pump power for the two corner sites ($C1$ and $C2$). For the following plots we fix the value of the pump at each site to the intensity indicated by the vertical gray lines. (b) Time dynamics of the corner sites, $C1$ and $C2$ in the presence of a continuous pump at both corners and pulse at  $C1$ (which is gradually turned on to avoid unwanted jumps in the initial stages). A pulse switches $C1$ from the lower to upper intensity state and then due to the parametric scattering,  $C2$ is also switched. We consider that the frequencies of both pumps and the central pulse frequency are same. All the blue curves correspond to $C1$ and red curves correspond to  $C2$. Parameters: $\Delta=-0.3$, $F_0=33$, $x_0=1$, $y_0=1$, $L=1$, $t_0=1283$, $\tau=15$.}
\label{fig:bistability}
\end{figure}

To show bistability we consider Eq.~(\ref{eq:GP1}) with the pump profile of the eigenstate $n=1251$ and $n=1249$ respectively, and  slowly vary the intensity of the pump in time (over $9000$ units $\sim 6000$ ps). The intensity corresponding to the corner sites $(C1 $ and $ C2)$ as a function of pump intensity is plotted in Fig.~\ref{fig:bistability}(a), where the characteristic hysteresis loops show that bistability is present (different corners are non-identical due to the lack of symmetry of the underlying lattice, which is why the hysteresis curves are slightly different). Gradually increasing the pump intensity to the level marked by the vertical grey lines in Fig.~\ref{fig:bistability}(a), allows each corner state to be initialized in its lower intensity state. 
Next, we apply a coherent Gaussian shaped pulse at corner $C1$ of the form
$
F = F_{0}\exp[-((x-x_{0})^2+(y-y_{0})^2)/L^{2}-(t-t_{0})^2/\tau^{2}-i\omega_{p}t],
$
where $F_0$ is the amplitude of the pulse which is launched at $(x_0,y_0)$, the coordinates of $C1$; $L$ and $\tau$ are the widths of the pulse in space and time. The time dynamics of both the corner sites in presence of the pumps and pulse is plotted in Fig.~\ref{fig:bistability}(b), which shows that the pulse switches $C1$ from its lower state to the upper state and then due to the parametric scattering, $C2$ also switches to the upper intensity state. The same switching does not occur in the absence of the pump at $C1$, that is, it is only when the first corner supports bistability that a binary signal can be transported and not just a direct effect of the applied pulse (see the SM).\par
All the calculations in this section were performed considering also an onsite disorder with uniform distribution and peak to peak magnitude of $0.03$. This physically corresponds to a disorder strength of 30 $\mu eV$ (for $J=1$ meV), which has been recorded experimentally \cite{Baboux_2016}.

{\textit{Advantage of topological corner states over regular square lattice}} ---
One could imagine that a similar scheme of coupling could occur in non-topological systems.
However we have found that for similar parameters, a regular square lattice operates much slower, not reaching a steady state even after $5000$ ps (see the SM). This is understandable from the fact that in the considered scheme, the linear decay is compensated by a nonresonant pump and the only dissipation present is the nonlinear decay ($\alpha_{NL} |\psi|^2 \psi$). Since the regular square lattice does not show any localized mode (all the modes are distributed over many sites), for a particular site the decay  $\alpha_{NL} |\psi|^2 \psi$  becomes very weak and consequently the corner site reaches a steady state very slowly. On the other hand, since the topological corner modes are perfectly localized at the corners, this problem does not arise. To solve this, we added some linear decay in the system which indeed made the system attain a steady state faster but we did not observe bistability in the case of a regular square lattice within the same window of pump intensity (or in fact a larger intensity window either). Bistability did occur in the lattice with corner modes in presence of linear loss. In this case polaritons are localized and experience a stronger nonlinear interaction than the case of delocalized polaritons in a regular square lattice that automatically spread out over a wider area. In other words, the advantage of the scheme with topological corner states over a regular lattice is faster operation with lower power.

{\textit{Conclusions---}} We considered the appearance of topologically protected corner states in a square array of coupled exciton-polariton micropillars. These systems can be distinguished from other topological photonic systems by the presence of significant nonlinearity arising from polariton-polariton scattering. Here we found that such processes allow corner states to be nonlinearly coupled to edge states, which can be further coupled to adjacent corner states. That is, in the nonlinear regime, edge states act as intermediaries between corners. It is generally speculated that topological modes can be relevant in information processing and we anticipate that the ability to couple multiple topological modes in a single system will be essential to such directions \cite{FlayacSavenko2013,BallariniGiorgi2013}. 

{\textit{Acknowledgments---}} We thank Sanjib Ghosh and Kevin Dini for helpful discussions and comments. The work was supported by the Ministry of Education, Singapore (grant no. M0E2017-T2-1-001).


\end{document}


\title{Supplemental Material for \\ Coupling between exciton polariton corner modes mediated through edge states}%

\author{R. Banerjee}\email[Corresponding author:~]{rimi001@e.ntu.edu.sg}
\author{S. Mandal}
\author{T.C.H. Liew}\email[Corresponding author:~]{tchliew@gmail.com}

\affiliation{Division of Physics and Applied Physics, School of Physical and Mathematical Sciences, Nanyang Technological University, Singapore 637371, Singapore}
\maketitle

{\textit{Continuous model}---}
Here we describe the scheme to realize the four different couplings used in our model and by solving the Schr\"odinger equation in a continuous space (without tight-binding approximation) of a coupled micropillars system we realize topologically protected corner states. We start by solving the time independent Schr\"odinger equation 
\begin{equation}
-\frac{\hbar^2}{2m}\nabla^2\psi(\mathbf{x})+V(\mathbf{x})\psi(\mathbf{x})=E\psi(\mathbf{x})
\label{seq1}
\end{equation} 

\begin{figure}[b]
\centering
\includegraphics[width=\linewidth]{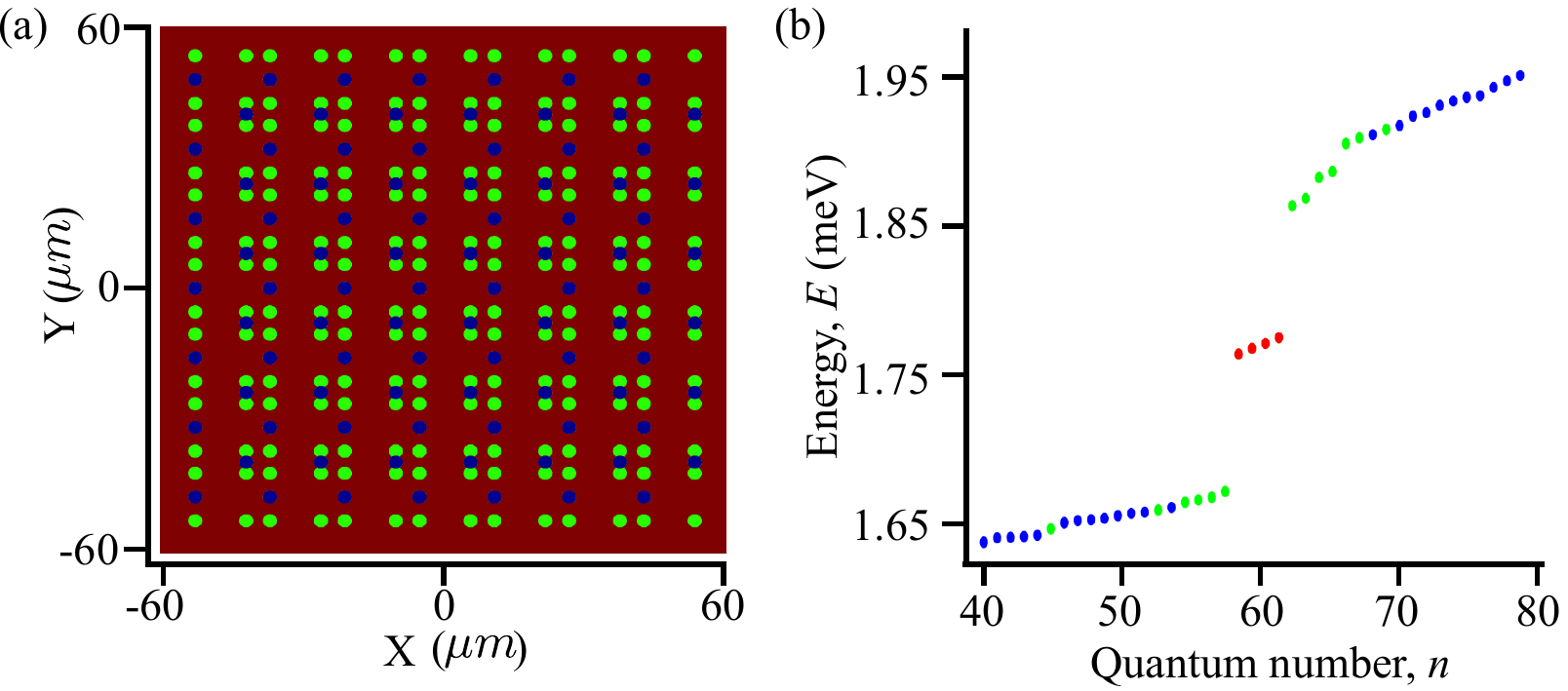}
\caption{(a) The potential profile of the system consisting of $14\times 14$ (main) micropillars. The green dots correspond to the main micropillars and the blue dots correspond to the auxiliary micropillars by which the negative hopping is realized. The center to center distance between the micropillars is varied to realize $J$ and $J^{\prime}$ couplings. (b) Plot of the energy eigenvalues of the system. The red dots correspond to the topologically protected corner states, the green dots correspond to the edge states and the blue dots correspond to the bulk states.}
\label{fig:potential}
\end{figure}
\begin{figure}[t]
\centering
\includegraphics[width=0.9\linewidth]{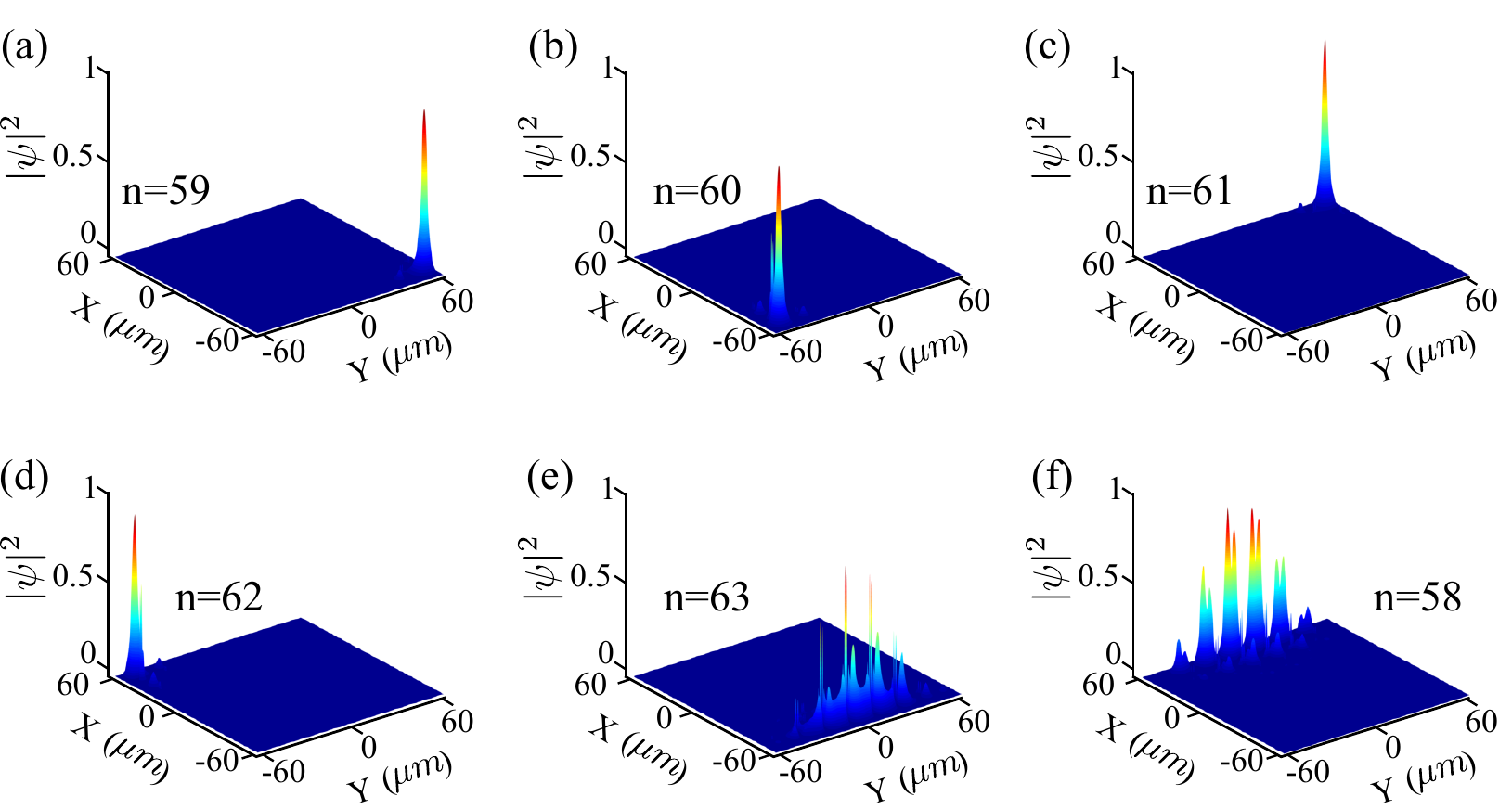}
\caption{Spatial profile of the corner states (a-d), and two (of many) edge states (e-f)obtained from the solution of Eq.~\ref{seq1}.}
\label{fig:spatial_profile}
\end{figure}


where $m$ is the polariton mass and $V(\mathbf{x})$ is the potential profile of the system described in Fig.~\ref{fig:potential}(a). In this scheme the micropillars are coupled evanescently in which the strength of the hopping constant can be controlled by changing the center to center distance between the micropillars. The micropillars are shown in green in Fig.~\ref{fig:potential}(a). The center to center distance between the micropillars is varied deliberately in order to realize two positive couplings in the corresponding tight-binding model of the main text, $J$ and $J^{\prime}$, where $J<J^{\prime}$. In order to realize the negative couplings ($-J$ and $-J^{\prime}$) we introduce another set of micropillars, which we call auxiliary pillars, in between the main micropillars. The auxiliary pillars are shown in blue in Fig.~\ref{fig:potential}(a). If the auxiliary pillars have  potential depth ($V_a$) greater  than that of the main micropillrs ($V_0$) then the negative hopping between the main micropillars can be realized \cite{Keil_2016}. In Fig.~\ref{fig:potential}(b) the eigenmodes corresponding to a $14\times 14$ system size are plotted. By proper choice of parameters, topologically protected corner modes can be realized. The spatial profile of the corner modes and edge modes are shown in Fig.~\ref{fig:spatial_profile}. The diameter of all the micropillars was taken to be 3 $\mu m$ \cite{Galbiati_2012}, and the center to center distances between the micropillars were $10.8~\mu m$ and $5.04~\mu m$ to realize $J$ and $J^{\prime}$ couplings, respectively. The auxiliary micropillars are placed exactly at the middle of the two main micropillars whenever the negative hopping between the main micropillars is needed. The potential depth of the main micropillars is $V_0=1.8$ meV and that of the auxiliary micropillars is $V_a=50.6$ meV. The mass of the polaritons $m=10^{-5}m_e$, where $m_e$ is the free electron mass. The obtained band gap corresponding to the above mentioned  parameters is around $0.2$ meV, which is higher than the typical disorder strength in the exciton polariton systems \cite{Baboux_2016}.\\

{\textit{Spatial profile of the instable modes in Fig.~2(b) in the main text}---}
\begin{figure}[b]
\centering
\includegraphics[width=0.85\linewidth]{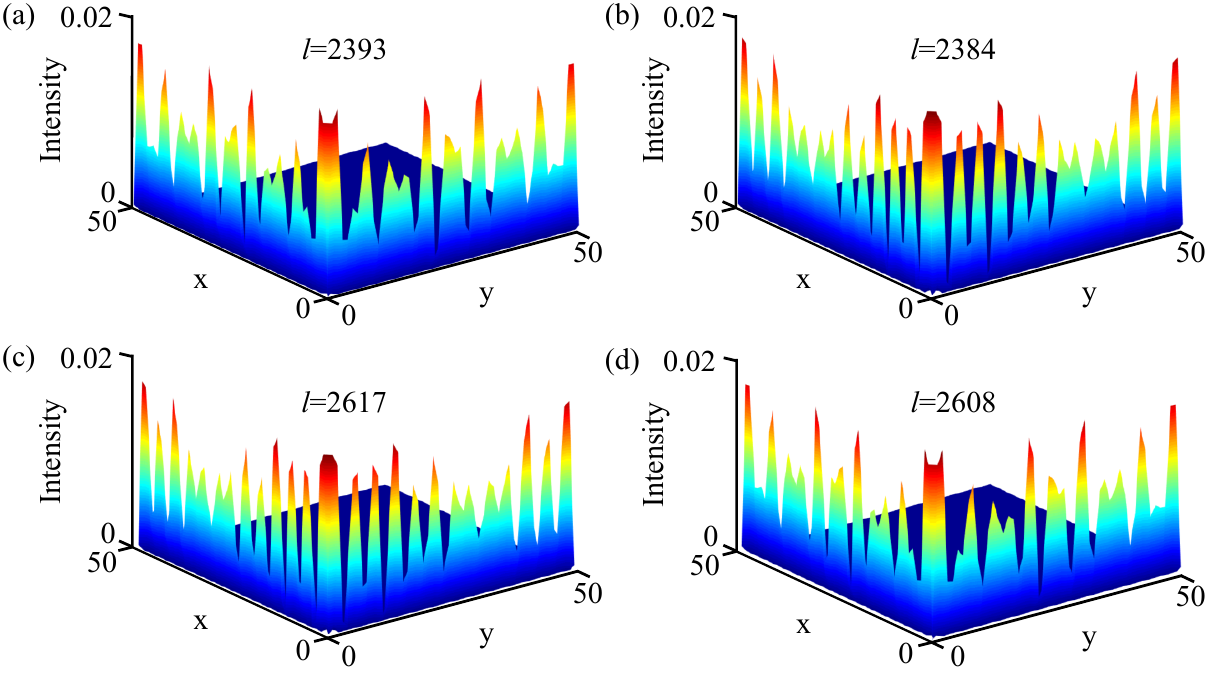}
\caption{Spatial profile of the intensity corresponding to the states whose Im($\omega$)$>0$ in Fig.~2(b) in the main text.}
\label{fig:spatial_profile_fig2edge}
\end{figure}
In Fig.~\ref{fig:spatial_profile_fig2edge} we plot the spatial profile of the intensity of the modes corresponding to  Im$(\omega)>0$ in Fig.~2(b) in the main text. The figure shows that indeed they are edge states. \\

\begin{figure}[htbp]
\centering
\includegraphics[width=\linewidth]{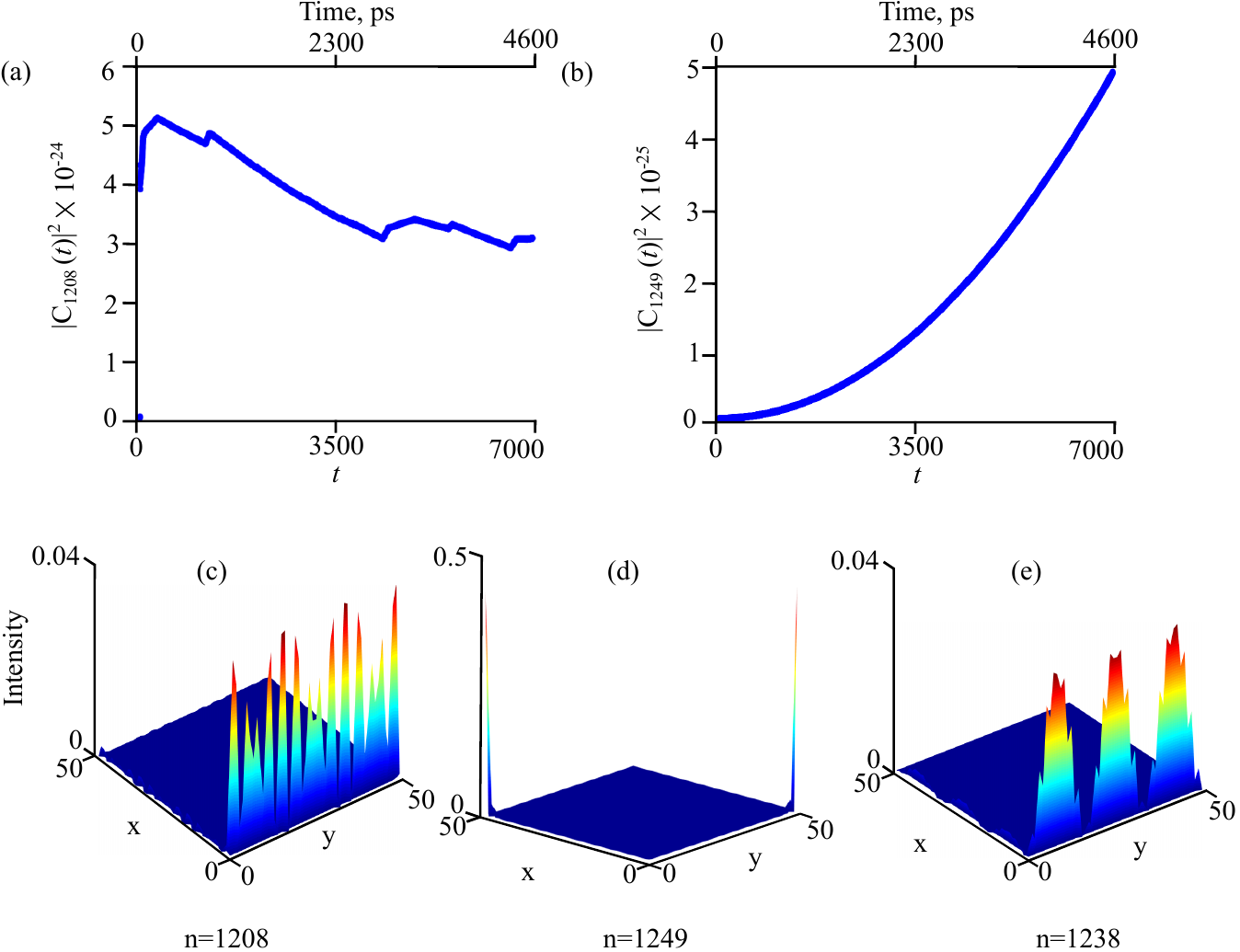}
\caption{ (a-b) Overlap $|C_{n}(t)|^2$ as a function of time corresponding to the same edge $n=1208$ and adjacent corner $n=1249$ in Fig.~3 (in the main text) but with no nonlinear terms.  (c-e) Spatial profile corresponding to the edge state $n=1208$, adjacent corner $n=1249$, and another edge state $n=1238$ respectively. }
\label{fig:fig3_noNONL_profile}
\end{figure}

{\textit{Overlap as a function of time in the absence of any nonlinear terms}---}To show that the parametric instability arises purely due to the nonlinearity in the system we evolve Eq.~(2) in the main text in time, from the initial condition $\psi_i=0$, in the absence of the nonlinear terms. We plot the overlap $|C_n(t)|^2$ corresponding to the edge state $n=1208$, and  for the adjacent corner state $n=1249$ in Figs.~S4(a-b). The order of the overlap implies that without nonlinearity, the coupling of the excited corner state to the edge states or its adjacent corner state is insignificant. In Figs.~\ref{fig:fig3_noNONL_profile} (c-e) we plot the spatial profile of the states corresponding to $n=1208$, $n=1249$, and $n=1238$, respectively, which shows that they correspond to an edge state, adjacent corner state, and another edge state. 

\begin{figure}[H]
\centering
\includegraphics[width=0.5\linewidth]{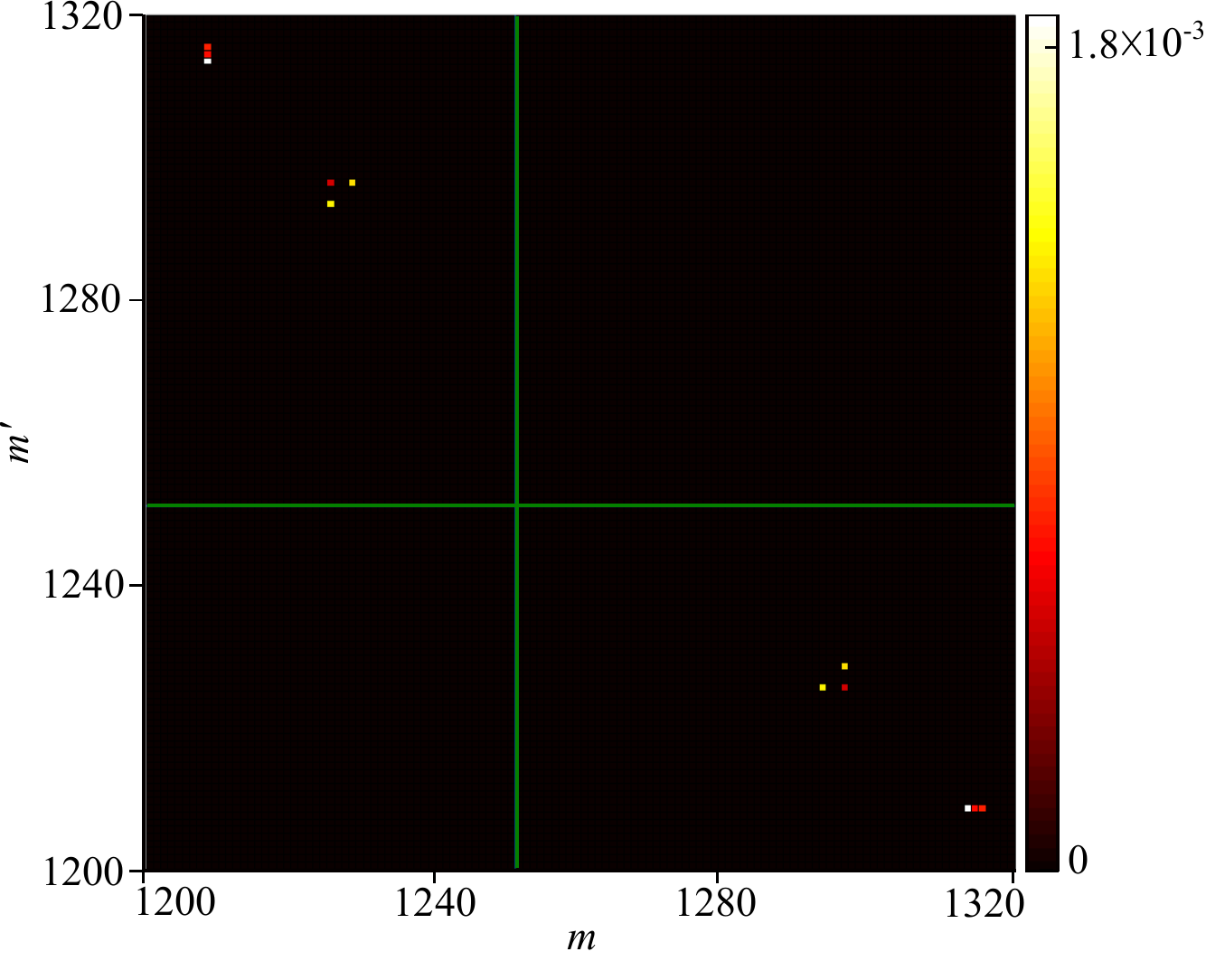}
\caption{Zoomed view of the positive imaginary value of the energy of the Bogoliubov spectrum in the  eigenstate basis calculated from Eq.~(\ref{eq:9}). The green lines corresponds to $m=c$ and $m'=c$. The figure shows that there is no instability for $m$ or $m'$ corresponding to any corner states, which proves that there is no direct coupling between two corners. The instability points correspond to the edge states showing that they are responsible for intermediating the coupling between adjacent corner states. }
\label{fig:m_mp}
\end{figure}

\begin{figure}[H]
\centering
\includegraphics[width=0.8\linewidth]{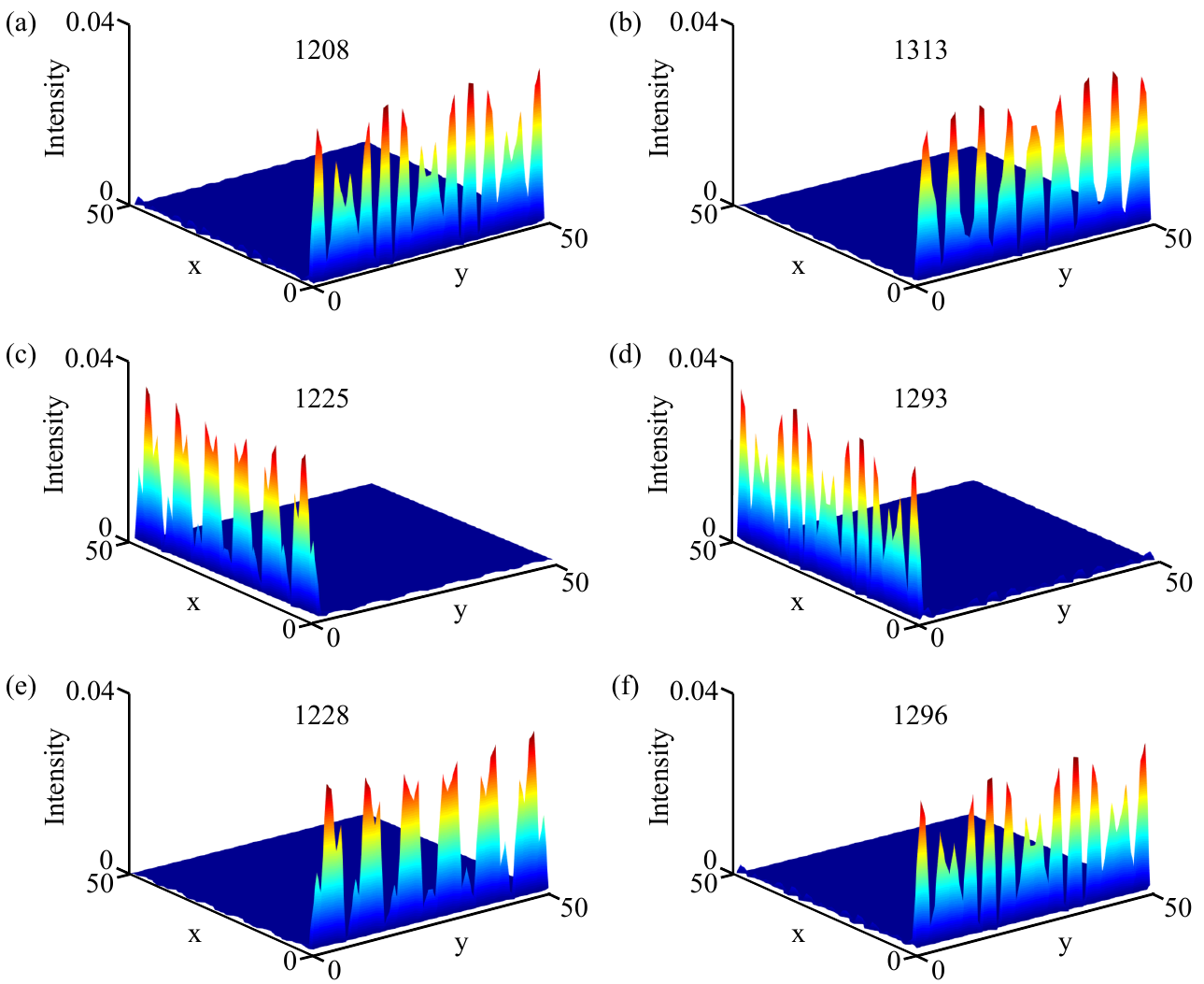}
\caption{Spatial profile corresponding to modes $m$ or $m'$ whose Im($\omega$)$>0$ in Fig.~\ref{fig:m_mp}. }
\label{fig:fig4_realspace_profile}
\end{figure}

{\textit{Proof of no direct coupling between corner states}---}
To show that the nonlinear terms do not directly couple the corner states, we switch to the eigenstate basis, using the transformation
\begin{align}
\tilde{\psi_n}=\sum_{i}(U^{-1})_{ni}\psi_{i}
\label{eq:5}
\end{align}
where $U$ is the unitary matrix that diagonalizes the linear problem. Eq.~(2) in the main text can be expressed in the eigen basis as
 \begin{align}
i\frac{\partial\tilde{\psi}_n}{\partial t}=E_{n}\tilde{\psi}_n+\sum_{ n' n'' n'''}\alpha_{n n' n'' n'''}\tilde{\psi}_{n'}^{*}\tilde{\psi}_{n''}\tilde{\psi}_{n'''}+\tilde{F}_{n}
\label{eq:6}
\end{align}
where $\alpha_{n n' n'' n'''}=\left(1-i\alpha_{NL}\right)\sum_{i}U^{-1}_{ni} U^{*}_{in'}U_{in''}U_{in'''}$ represents the effective polariton-polariton interaction and nonlinear loss, which is no longer a local term (instead it is $n$ dependent). Here $\tilde{F}_{n}=\sum_{i}(U^{-1})_{ni}F_{i}$ is the coherent excitation term in the eigen basis. We recall that the coherent pump $F$ has the profile matching the corner state $n=1251$, which we define as $n=c$. Consequently, $\tilde{F}_{n}=0$ for $n\neq c$, and the second term in the r.h.s of Eq.~(\ref{eq:6}) will only survive if $n'=n''=n'''=c$. The dynamics of the system is then governed by

\begin{align}
i\frac{\partial\tilde{\psi_c}}{\partial t}=E_{c}\tilde{\psi_c}+\alpha_{c c c c}|\tilde\psi_{c}|^{2}\tilde\psi_{c}+\tilde{F_{c}}
\label{eq:7}
\end{align}
Here $E_{c}$ is the energy detuning between the eigen-value corresponding to the corner and laser. Once the stationary state ($\tilde{\psi}_{0,c}$)  from Eq.~(\ref{eq:7}) is obtained, we can define the linear Bogoliubov fluctuations of the form
\begin{align}
\tilde{\psi_n}=\tilde{\psi}_{0,c} \delta_{n,c}+\tilde{u}_{n}\delta_{n,m}e^{-i\omega t}+\tilde{v}_{n}^{*}\delta_{n,m'}e^{i\omega^{*} t}
\label{eq:8}
\end{align} 
which demonstrate that a pair of  polaritons from the stationary state can scatter to states $n=m$ and $n=m'$ conserving energy. Inserting  Eq.~(\ref{eq:8}) into Eq.~(\ref{eq:6}):
\begin{align}
\omega\tilde{u}_m&=\left(E_{m}+(\alpha_{m c c m}+\alpha_{m c m c})|\tilde{\psi}_{0,c}|^{2}\right)\tilde{u}_m+\alpha_{m m' c c}\tilde{\psi}_{0,c}^{2} \tilde{v}_{m'}\notag\\
\omega\tilde{v}_m'&=\left(-E^{*}_{m'}-(\alpha_{m' c c m'}+\alpha_{m' c m' c})^{*}|\tilde{\psi}_{0,c}|^{2}\right)\tilde{v}_m'-{\alpha}^{*}_{m' m c c}(\tilde{\psi}_{0,c}^{2})^{*} \tilde{u}_{m}
\label{eq:9}
\end{align} 
Physically Eq.~(\ref{eq:9}) and Eq.~(4) in the main text are the same although written in different bases. Solving the eigenvalues for all possible pairs of $m,m'$ we obtain Fig.~\ref{fig:m_mp} where only the positive imaginary values are plotted. We see that there is no instability for $m$ or $m'$ corresponding to the corner states, which proves that the intensity dynamics plotted for the adjacent corner in Fig~3(c) in the main text is not due to the  direct coupling between the corners. We can conclude that polaritons from the pumped corner first couple to the edge states, then the polaritons from the edge states couple to the adjacent corner resulting in a nonzero intensity at the adjacent corner.  One of the indices corresponding to the maximum instability is  $m=1208$, whose overlap is plotted in Fig~3(b) in the main text. The state in Fig.~3(d) (in the main text) does not appear in Fig.~\ref{fig:m_mp} indicating that its significant intensity is due to higher order scattering processes not taken into account in the linear Bogoliubov theory. We checked that all the states having positive imaginary values in Fig.~\ref{fig:m_mp} are edge states and the intense modes (white and yellow ones) are plotted in Fig.~\ref{fig:fig4_realspace_profile}. Their nonzero overlap as calculated from Eq. (5) in the main text, indicating the coupling with the excited corner state. 

\begin{figure}[H]
\centering
\includegraphics[width=0.4\linewidth]{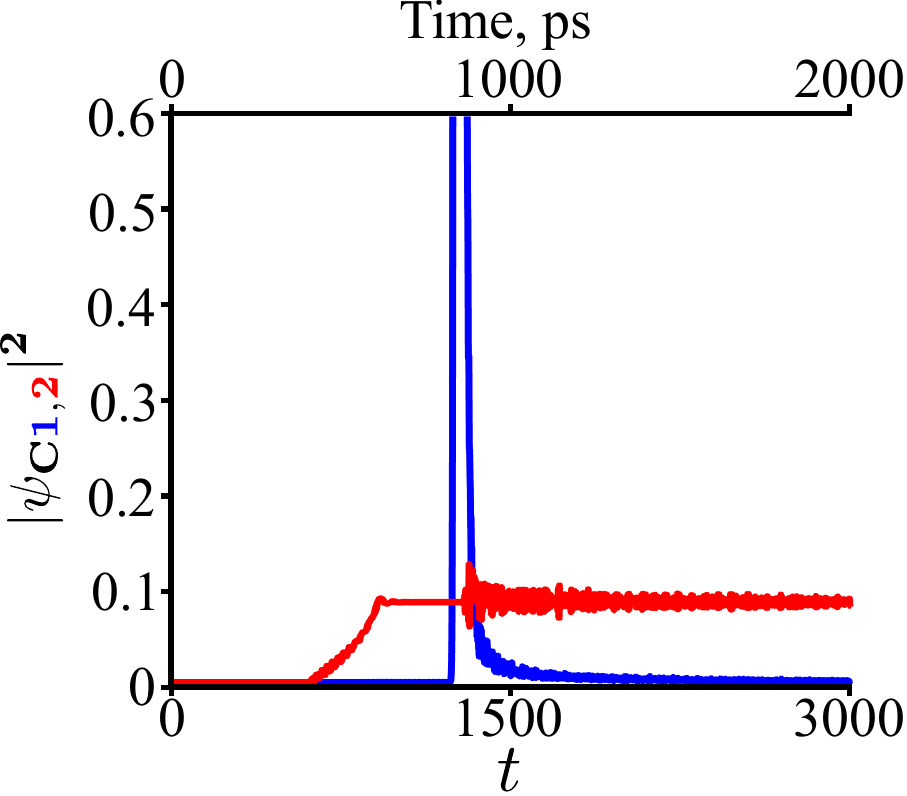}
\caption{Time dynamics of the two corner sites $C1$ and $C2$ in the presence of the same pulse at $C1$ and the continuous pump at $C2$, that is, same as in 4(b) in the main text, but without the continuous pump at $C1$. The intensity in $C1$ is insignificant because there is no continuous pump but $C2$ is still in the lower state. It implies that the pulse alone cannot switch the adjacent corner. The blue curve correspond to $C1$ and red curve correspond to  $C2$. }
\label{fig:pulseC1_profile}
\end{figure}

{\textit{Repeating 4(b) in the main text without the continuous pump at $C1$}---}
To show that the switching at corner $C2$  is due to the parametric scattering and the pulse at $C1$ can not alone induce switching at $C2$, we repeat 4(b) but without the continuous pump at $C1$. The time dynamics plotted in Fig.~\ref{fig:pulseC1_profile} shows that in this case the site $C2$ stays at the lower intensity state. 

\begin{figure}[htbp]
\centering
\includegraphics[width=0.8\linewidth]{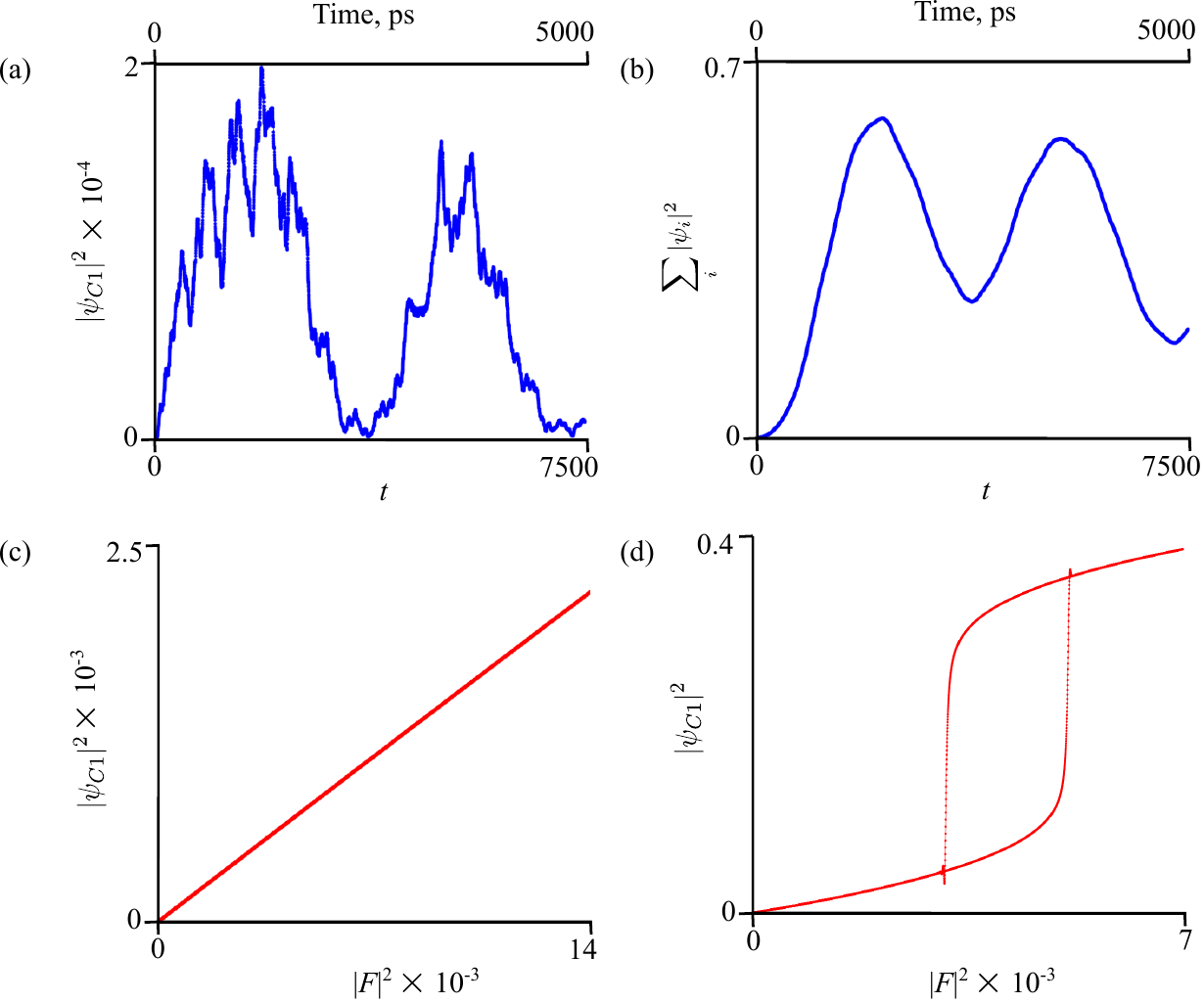}
\caption{(a-b)Time dynamics of the same corner site $C1$ and the total intensity of the regular square lattice for the pump of intensity $1\times10^{-3}$ respectively. Parameter: $\Delta=-0.3$. (c) Plot of the population of $C1$ of the regular square lattice as a function of pump intensity. (d) Hysteresis curve of the corner modes in presence of linear decay. In (c) and (d) a supplementary dissipation $\gamma=0.05$ is included according to Eq.~\ref{eq:GP1line} to ensure that the regular square lattice reaches a stationary state in a reasonable time.}
\label{fig:fig6_square}
\end{figure}

{\textit{Comparison of bistability in topological corner and regular square lattices.}---}
Here we show the advantage of information transfer with topological states over trivial states. We consider the same parameters and the same pump profile as the ones used in Fig.~4 in the main text, for a simple square lattice with the same average coupling strength between sites. We observe that under this condition, the square lattice does not reach a steady state even with 5000 ps (see Figs.~\ref{fig:fig6_square}(a-b), where we plot the time dynamics of the same corner site ($C1$) and the whole system, respectively). In contrast, the topological corner modes reach their steady state around 700 ps (see the blue curve before the pulse was launched in Fig.~4(b) in the main text). It happens due to the fact that in the considered model the only dissipation present is the nonlinear decay ($\alpha_{NL} |\psi_i|^2 \psi_i $), which should grow to compensate the pump $P$ to form a steady state. For the square lattice, since there is no localized mode, for a particular site, the nonlinear decay term is very weak and as a result it will reach steady state very slowly. On the other hand, the topological corner modes are well localized and as a result this problem does not arise. To resolve this, we add some linear decay ($\gamma$) in the system, which modifies Eq.~2 in the main text as,

\begin{align}
i\frac{\partial\psi_{i}}{\partial t}=\left(\Delta-\frac{i\Gamma}{2}+iP-i\gamma\right)\psi_{i}+J_{ij}\sum_{\langle j\rangle}\psi_{j}+\left(1-i\alpha_{NL}\right)|\psi_{i}|^2\psi_{i}+F_i
\label{eq:GP1line}
\end{align}

It indeed helps the system to reach the steady state quickly, but we do not observe bistability for the regular square lattice within the same window of pump intensity (or in fact a larger intensity window either, see Fig.~\ref{fig:fig6_square}(c)). But we still observe bistability for the corner modes even with the linear decay, shown in Fig.~\ref{fig:fig6_square}(d). In this case polaritons are localized and experience a stronger nonlinear interaction than the case of delocalized polaritons in a square lattice that automatically spread out over a wider area.

In other words, the advantage of the scheme with topological corner states over a regular lattice is that it operates faster and at lower power.    

\pagebreak